\documentclass[10pt,journal,compsoc]{IEEEtran}
\usepackage{graphicx}
\usepackage{hyperref}
\usepackage{comment}
\usepackage{xcolor}

\ifCLASSOPTIONcompsoc
  \usepackage[nocompress]{cite}
\else
  \usepackage{cite}
\fi
\ifCLASSINFOpdf
\else
\fi
\begin{document}
%\title{Prompt Sapper: LLM-Empowered AI Chain Methodology and Production Platform}
\title{Prompt Sapper: LLM-Empowered Software Engineering Infrastructure for AI-Native Services}

\author{Zhenchang~Xing,~\IEEEmembership{Member,~IEEE,}
        Qing~Huang,~\IEEEmembership{Member,~IEEE,}
        Yu~Cheng,% <-this % stops a space
        ~Liming~Zhu,~\IEEEmembership{Senior Member,~IEEE,}
        Qinghua~Lu,~\IEEEmembership{Member,~IEEE,}
        and~Xiwei~Xu,~\IEEEmembership{Member,~IEEE,}% <-this % stops a space      
\IEEEcompsocitemizethanks{\IEEEcompsocthanksitem Z. Xing is with CSIRO's Data61, Australia.\protect\\
E-mail: zhenchang.xing@data61.csiro.au
\IEEEcompsocthanksitem L. Zhu, Q. Lu and X. Xu are with CSIRO's Data61, Australia.
\IEEEcompsocthanksitem Q. Huang and Y. Cheng are with Jiangxi Normal University, China. Q. Huang is the corresponding author.}% <-this % stops an unwanted space
\thanks{Manuscript received XXX XX, XXXX; revised XXX XX, XXXX.}}


\markboth{Journal of \LaTeX\ Class Files,~Vol.~14, No.~8, August~2015}%
{Shell \MakeLowercase{\textit{et al.}}: Bare Demo of IEEEtran.cls for Computer Society Journals}



\IEEEtitleabstractindextext{%
\begin{abstract}
Foundation models, such as GPT-4, DALL-E have brought unprecedented AI ``operating system'' effect and new forms of human-AI interaction, sparking a wave of innovation in AI-native services, where natural language prompts serve as executable ``code'' directly (prompt as executable code), eliminating the need for programming language as an intermediary and opening up the door to personal AI.
Prompt Sapper has emerged in response, committed to support the development of AI-native services by AI chain engineering. 
It creates a large language model (LLM) empowered software engineering infrastructure for authoring AI chains through human-AI collaborative intelligence, unleashing the AI innovation potential of every individual, and forging a future where everyone can be a master of AI innovation.
This article will introduce the R\&D motivation behind Prompt Sapper, along with its corresponding AI chain engineering methodology and technical practices.
\end{abstract}

}

\maketitle

\IEEEdisplaynontitleabstractindextext

% \fi
\IEEEpeerreviewmaketitle

\IEEEraisesectionheading{\section{Introduction}\label{sec:introduction}}

\begin{quote}
\textit{"The greatest danger in times of turbulence is not the turbulence - it is to act with yesterday's logic."}\\
\hspace*{\fill} --- Peter Drucker
\end{quote}

\begin{figure*}
    \centering
    \includegraphics[width=\textwidth]{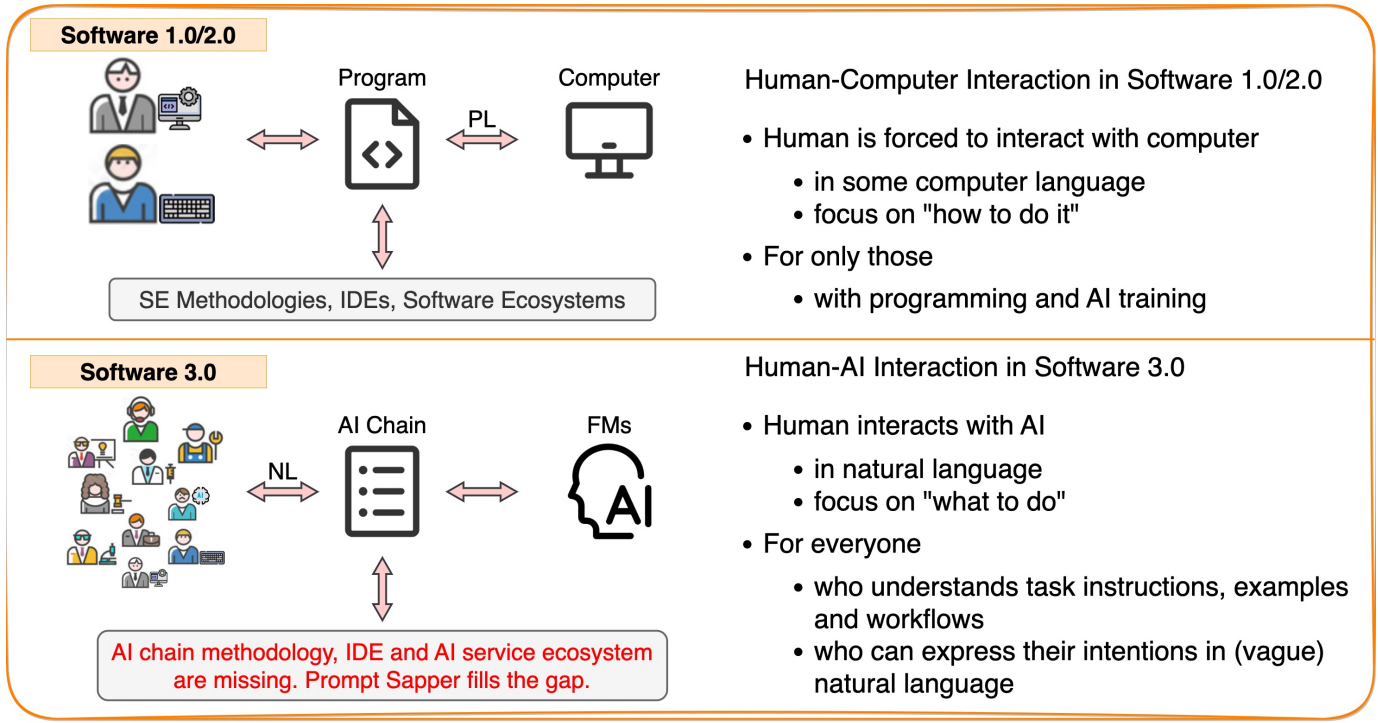}
    \caption{Foundation Models Unlock Software 3.0 (The most essential characteristic of Software 3.0 is ``prompt as code'')}
    \label{fig:software3.0}
\end{figure*}

Just as 40 years ago, personal computers (PCs) has brought about the transformation of information age; 
15 years ago, the proliferation of smartphones further drove the development of mobile computing and applications, making our daily lives and work more convenient and efficient. 
Now, the rapid development of artificial intelligence (AI) technology, represented by foundation models~\cite{bommasani2021opportunities} such as the well-known GPT-4~\cite{openai2023gpt4} and DALL-E~\cite{ramesh2021zero} , is bringing about the long-awaited AI platformization effect and a new trend of technological revolution.

In the Project AI 2.0~\cite{ai-2-0}, Kai-Fu Lee believes that the current popular chatbots and text-to-image generation are just the tip of the iceberg in this revolution.
As illustrated in Figure~\ref{fig:software3.0}, foundation models unlock Software 3.0. 
The most essential characteristic of Software 3.0 lies in ``prompt as code'', that is, we can interact with AI and instruct AI with our intent in natural language ``code'' (i.e., prompts) directly to create AI-native services on top of foundation models.
In Software 1.0, the code is written in a programming language by developers. 
In Software 2.0~\cite{AndrejSoftware2.0}, the network behavior is learned from data, but the neural network code is still written in a programming language by developers.

Foundational models would go beyond making software 1.0/2.0 development more productive, for example by Github Copilot (i.e., prompt to code).
They are changing not only who can create software but also what types of software can be created~\cite{wu2022promptchainer}.
Software 3.0 frees people from traditional programming and AI training, and allows (non-technical) individuals to create, customize and compose AI services by describing ``what to do'' (e.g., task workflow and data characteristics), and designing human-AI interaction through natural language, rather than focusing on ``how to do'' as in Software 1.0/2.0, i.e., data structures, algorithms, API calls, data labeling, and model training.

The bottleneck of \textit{personal AI} (analogous to PC 40 years ago) is rapidly opening up thanks to foundation models.
If we regard the foundation models as AI ``operating system'', the generative capabilities invoked by prompt-as-code are the ``API'' to the AI ``operating system'', but they are not AI-native services running on the AI ``operating system''.
We are in the stage of transforming the capabilities of foundation models into practical AI services, which presents two opportunities: 1) allowing everyone (not just AI or software engineers) to create personalized AI services, and 2) enabling people to share and hire AI services. 
Andrej Karpathy envisioned that this natural language programming paradigm could potentially expand the number of ``programmers'' to 1.5 billion~\cite{Andrej2023} .

%This will allow a broader audience to participate in the AI wave and benefit from it.
Despite the rapid emergence of AI services based on foundation models (see \href{https://gpt3demo.com/}{https://gpt3demo.com/}), their development still requires traditional software development methods, offsetting the AI democratization effect of foundation models, and thus many great ideas remain stagnant at the early ideation stage and fail to progress further. 
While ChatGPT demonstrates impressive capabilities, including generating and testing code, chat is not inherently an effective software development mechanism.
Furthermore, prompt-based AI services relies on prompt engineering magic which lacks transparency and is hard to control, test, reuse and maintain, akin to the state of ad-hoc coding before the advent of software engineering 55 years ago.

In response to the AI 2.0 and Software 3.0 revolution, what we need is not just flashy prompts and chatbot, but a transformative software engineering infrastructure that allows everyone to participate and truly benefit from AI.

\section{AI Chain Engineering}

\begin{figure*}
    \centering
    \includegraphics[width=\textwidth]{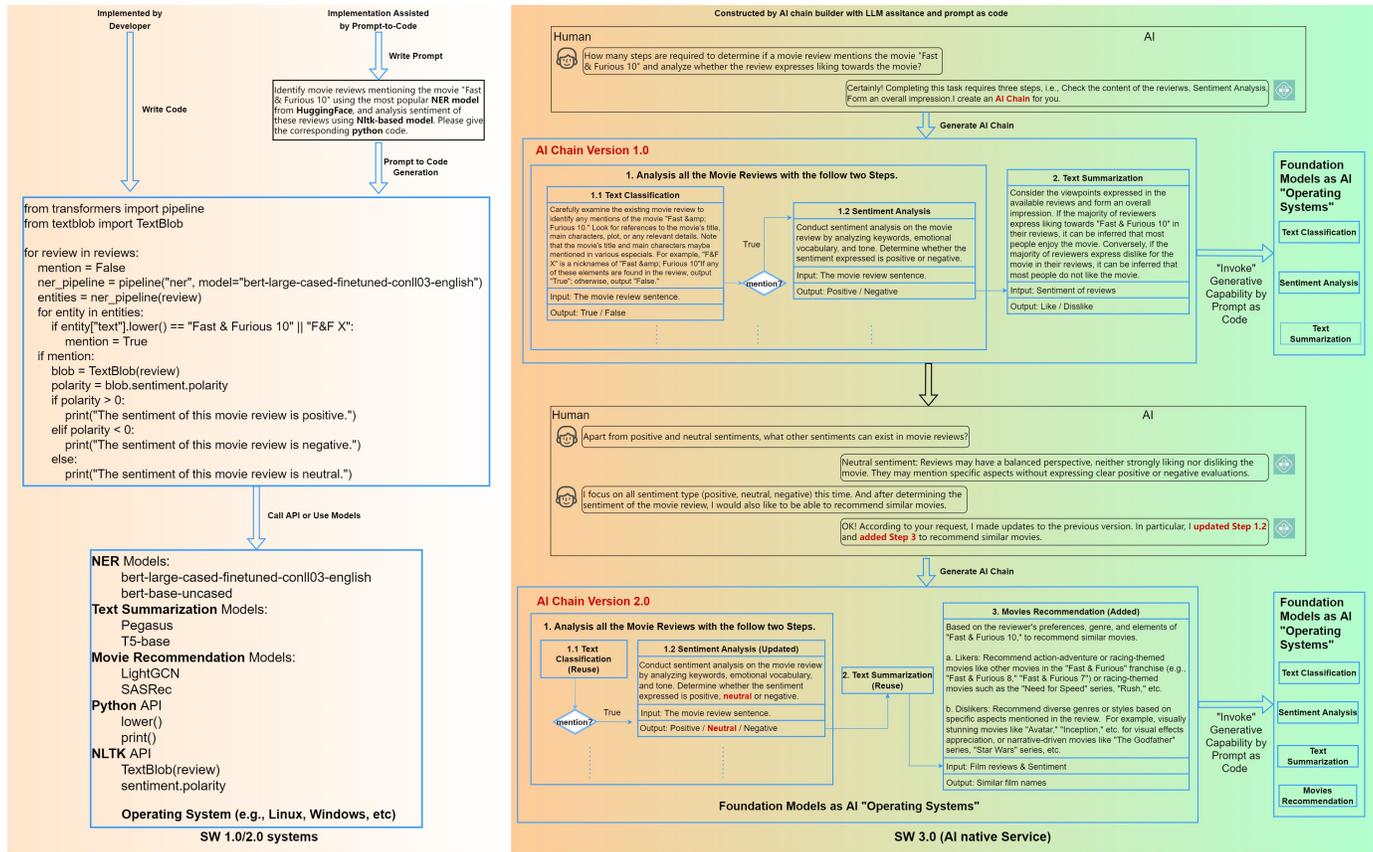}
    \caption{Building AI-Native Services through AI Chain Engineering, with LLM Assistance and Prompt-as-Code}
    \label{fig:ainativeservice}
\end{figure*}

Our vision is to reshape software landscape through generative AI.
To realize this vision, we propose \textit{AI chain engineering}\footnote{As an emergent paradigm, we focus on the development of AI-native services through AI chain engineering, and leave the maintenance and other DevOps activities as future work.}.
We regard foundation models as AI ``operating systems'' whose generative capabilities can be invoked by prompt-as-code, without requiring a programming language as an intermediary.
AI chains are a novel form of software product that assembles prompt calls to foundation models, together with calls to traditional AI models, external data or APIs (if needed), according to specific workflows, thereby delivering AI-native services.
AI chain functions as both a medium for human-AI collaboration, and as the tangible outcome of such collaboration.
They intuitively map to task workflow and are ``coded'' in natural language prompts, directly learnable and moldable by ordinary people (AI chain builders).

As illustrated in Figure~\ref{fig:ainativeservice}, the AI chain builders do not need to switch to ``programmer mindset'' as in Software 1.0/2.0 but stay within their task contexts for building AI-native services.
In Software 1.0/2.0, the user requires a good knowledge of the solution space, for example, different types of models and APIs and how to use them in source code.
AI assistance (e.g., Github Copilot) could lower this technical barrier, but the user still needs to know some technical concepts and tools (e.g., NER, HuggingFace, NLTK) in order to write the effective prompts for code generation.
In Software 3.0, the LLM assists the user in acquiring task knowledge (e.g., sentiment ratings), analyze requirements, decompose tasks, and generate and evolve AI chains.
AI chains consist of AI-native workers that directly invoke emergent capabilities of foundation models by natural language prompts (i.e., prompt as code), without the need for programming language as an intermediary.
AI chain workers uphold well-established software engineering values, such as modularity, extensiblity, reusability.

\begin{figure*}
    \centering
    \includegraphics[width=\textwidth]{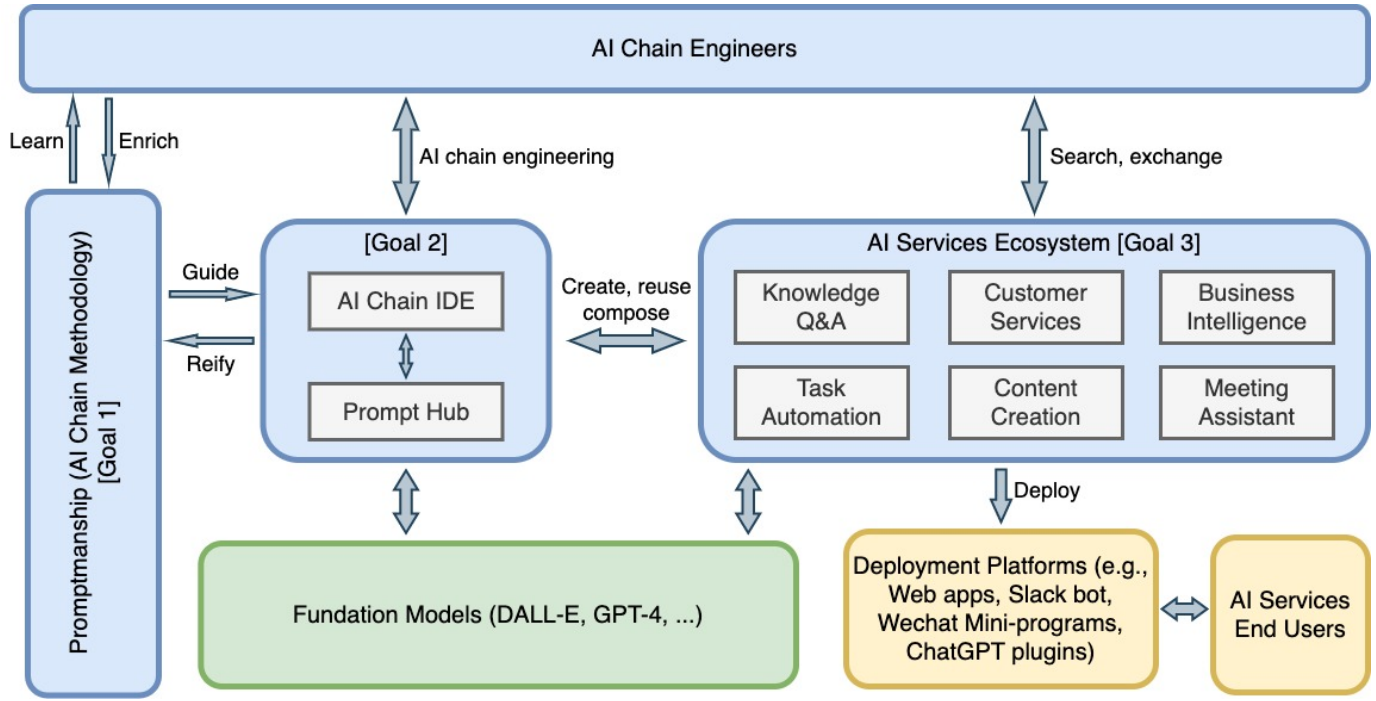}
    \caption{Software Engineering for AI Chain: Vision and Goals}
    \label{fig:se4aichain}
\end{figure*}

As shown in Figure~\ref{fig:se4aichain}, we are committed to develop a systematic AI chain methodology (Goal 1: Promptmanship), the corresponding AI-native development and deployment environment (Goal 2: AI chain IDE), and the AI services marketplace and ecosystem (Goal 3). 

First, prompt-as-code does not imply that writing prompts is the only task AI-native service development entails.
We aim to summarize the best practices of prompt engineering and place them within the broad context of software engineering, complementing important software engineering methods that have been overlooked (such as software processes, system design, testing). 
This creates an AI chain methodology that upholds well-established software engineering principles and values to improve the transparency, controllablity, testability and maintainability of AI services built on top of foundation models.

Second, chatbot, regardless of the capabilities of the underlying LLM, is a only ``command-line console'' to the AI ``operating system''. 
We aim to develop an AI-native integrated development environment (IDE) that supports the whole process of AI chain development, from ideas to services. Through the tangible implementation of the AI chain methodology, AI chain IDE will support individuals to develop high-quality, foundation model-based AI services.
Leveraging natural language interface, LLM-empowered interactive guidance and automatic AI chain generation, this IDE lowers the technical barrier for AI chain development and improve the development efficiency. 

Third, we aim to develop an AI services marketplace to promote the development of the AI services ecosystem. 
We envision this AI services marketplace would be different from code repositories like Github and data/model hubs like HuggingFace, as the AI-native services are primarily prompt driven and offer practical compositional AI capabilities due to the AI platformization brought by foundation models.
We propose to investigate responsible AI chain engineering methods and technologies to enhance the transparency, accountability, and security of AI services.

We envision that AI chain builders adhere to promptmanship, develop AI chains in the AI-native IDE, and share them in the marketplace.
These AI-native services can be deployed in various platforms to serve the end users through the chatbot interface or other GUIs.
Our vision is to establish an open, collaborative, secure, and sustainable AI chain ecosystem that provides support for digital transformation and upgrade across various industries. 
We want to empower individuals and businesses to unleash their creativity and intelligence in the new era of AI 2.0 and Software 3.0, allowing them to benefit from it and achieve the vision of symbiosis between humans and AI.

\section{Promptmanship}

\begin{figure*}
    \centering
    \includegraphics[width=\textwidth]{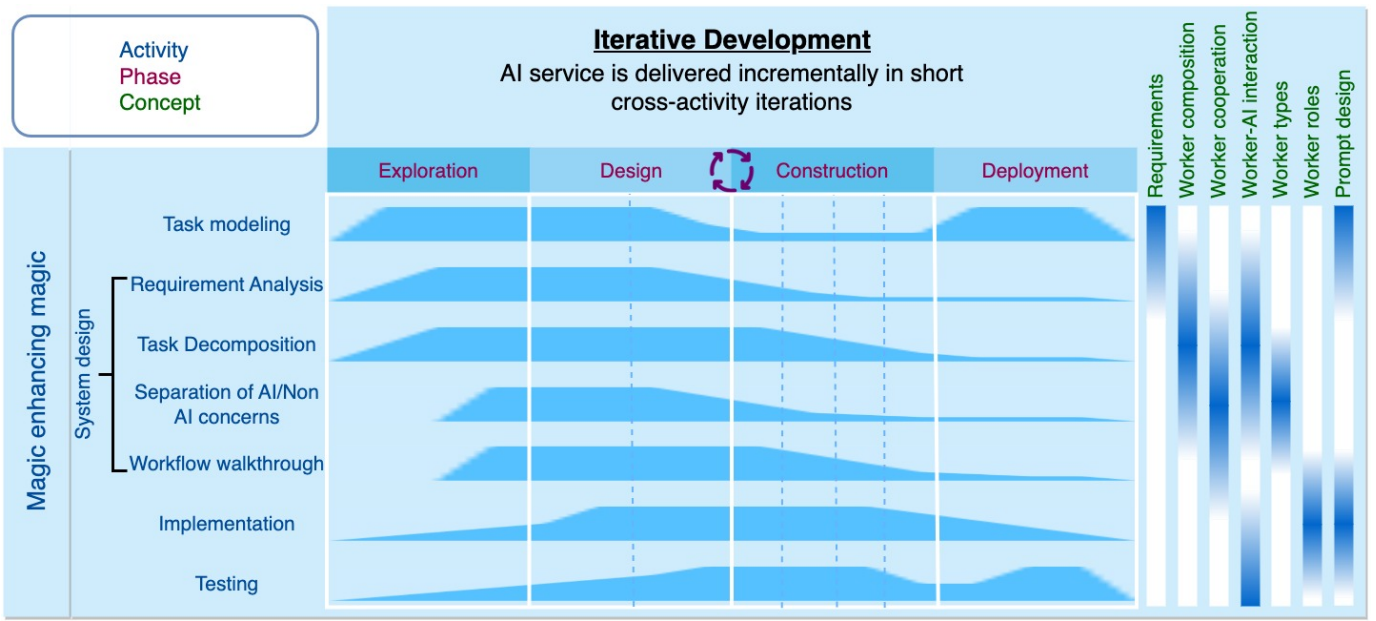}
    \caption{Promptmanship: AI Chain Process, Concepts, and Activities}
    \label{fig:promtpmanship}
\end{figure*}

Over the decades of its development, software engineering has accumulated a wealth of effective methods and practices, many of which can be applied to AI chain engineering. 
However, with the transformative shift in human-AI interaction brought about by foundation models, it is necessary to examine and adjust traditional software engineering methods and practices to accommodate the emerging human-centered natural language programming paradigm.

On one hand, LLMs encode vast amounts of knowledge and possess powerful conversational abilities. 
We can leverage them to assist AI chain engineers in acquiring domain knowledge, understanding problems, and gaining inspiration for problem-solving. 
At the same time, to mitigate the inherent challenges of errors and illusions in LLMs, we need to have a deep understanding of their capabilities (known as ``mechanical sympathy'') and employ effective prompt design patterns.
On the other hand, LLMs not only change who can develop AI services but also profoundly alter the types of AI services that can be developed. 
This requires a shift from code-centric development tools of the past to human-centric tools that enable (non-technical) individuals to focus on problem-solving and engage in intuitive analysis, design, construction, and evaluation of AI chains.

Guided by these design considerations, we propose an AI chain methodology (\href{https://www.aichain.online/public/content\%20pages/promptmanship.html}{promptmanship}) as shown in Figure~\ref{fig:promtpmanship}. 
We define the functional units of AI chains as ``workers''. 
Promptmanship encompass not only traditional software concepts such as requirements, object composition and collaboration, and object roles but also specific AI chain concepts. 
For instance, we differentiate three types of workers (corresponding to three software paradigms: Software 1.0/2.0/3.0\footnote{We believe Software 3.0 will not cover 100\% the problem space where Software 1.0 and Software 2.0 workers can perform more effectively and economically.
For example, Software 1.0 can process deterministic logic precisely, and Software 2.0 will be efficient and economic for the text classification task.}), four levels of worker-AI interaction modes based on increasing reasoning capabilities, nine worker stereotypes for input transformation, task processing and output validation, and 12 well-adopted prompting best practices (e.g., prompt caveats, aspects and decorators).

We view AI chain engineering as an iterative rapid prototyping process with many-step collaboration between human and AI.
This process consisting of four iterative stages: Explore, Design, Build, and Deploy. 
Each stage involves concurrent activities (inspired by the Rational Unified Process~\cite{kruchten2004rational}), including task modeling, system design (requirements analysis, task decomposition, AI/non-AI separation of concerns, workflow rehearsals), AI chain implementation, and testing.
However, each stage has a different emphasis (as depicted by the corresponding bar heights). 
Different activities generate or refine different AI chain concepts (shown within the blue intervals below the concepts). 
Throughout all activities, we propose a ``magic enhances magic'' activity that leverages the knowledge and conversational abilities of large language models to assist AI chain engineers in acquiring task knowledge, gathering and analyzing requirements, and gaining an understanding of model capabilities (i.e., mechanical sympathy).

At the end of an iteration, an AI chain will be constructed suitable for production.
It is fundamentally modular, as opposed to epic prompts that ask the LLM to do the whole thing all at once (so called \href{https://explosion.ai/blog/against-llm-maximalism}{LLM maximalism} by Matthew Honnibal).
The whole AI chain, as well as its individual workers, have been tested and optimized, and errors can be attributed to different workers and fixed.
When requirements change or get extended, new workers can be added or existing ones can be updated or replaced.
Workers can also be reused across different services.

\section{Sapper IDE}

Our Sapper IDE differs from existing code-centric development tools, because AI chains will be developed by many non-technical individuals. 
In the era of Software 3.0, these individuals possess vast demands and domain knowledge, but they lack proficiency in computer science, software development, and AI background knowledge. 
Consequently, they face challenges in effectively expressing their requirements and knowledge to AI and leveraging foundation models to create the desired AI services.

Therefore, our foremost design principle is ``human-centric'', which manifests in three aspects:
First, we reify the promptmanship (AI chain methodology) into the AI chain IDE, enabling anyone to effectively apply best AI chain practices and methods.
Second, we leverage the knowledge and conversational capabilities of LLMs to develop AI co-pilots (acting as virtual product manager, system designer and tester) that provide the whole-process AI chain development support for non-technical individuals.
Third, we offer a no-code AI chain analysis, design, construction, testing, and deployment process, making it easy for anyone to transform their ideas into AI services.

Based on these design principles, we have developed the Sapper IDE\footnote{Visit \href{https://promptsapper.tech}{https://promptsapper.tech} to try our Sapper IDE.}, an AI-native platform that fully supports the AI chain process, concepts and activities proposed in our promptmanship.
The Sapper IDE adopts a human-AI teamwork strategy, where humans, assisted by AI, mainly handle the two ends (expressing requirements, acceptance tests), while AI operates in the middle (generating and testing AI chains).
As such, different from the current IDE designed around code editor, the emphasis of Sapper IDE is on the LLM-empowered exploration and design view that iteratively scaffold requirement analysis and system design and on iterative AI chain testing and debugging, rather than on the visual programming as in~\cite{wu2022promptchainer}.

%Through natural language interface, interactive guidance and automatic AI chain generation, Sapper IDE lowers the technical barrier for AI chain development and improve the development efficiency. 
The Sapper IDE is built on a suite of LLM-empowered AI co-pilots for AI chain analysis, design, construction and testing.
This is, it is an LLM-empowered software engineering framework for AI chain, i.e., AI4SE4AI.
The Sapper IDE can be seen as an ``incubator for AI services'' as it leverages the power of foundation models to create more AI services through the deep collaboration between human and AI. 
It not only fulfills the users' needs for AI-native services but also inspires them to explore more possibilities and help them create even better AI services. 
%We believe this will be an era of limitless innovation, and our Sapper IDE will serve as the "tool" for various industries to unlock this infinite potential!

\section{AI Services Marketplace}

To demonstrate the applicability of our promptmanship and Sapper IDE, we have built a set of \href{https://www.aichain.online/public/content\%20pages/showcases/showcases.html}{AI service showcases} based on foundation models, including writing assistants, poem to paint generation, mental health assistant, interview training, programming assistant, quiz maker.
We are actively building an \href{https://aichain.store}{AI service marketplace} and will allow users to directly share their AI chain projects from the Sapper IDE to the marketplace. 
We anticipate that the community will contribute a multitude of excellent personalized or vertical market AI services, fostering a rich and diverse AI services ecosystem centered around foundational models. 
Open-source AI chain projects will provide ample learning materials and inspiration to onboard beginners.

%This will support digital and intelligent transformation across various industries, creating extraordinary value. 

This AI services marketplace would be different from code repositories like Github and data/model hubs like HuggingFace, as AI-native services are primarily prompt driven and offer the compositional AI capabilities due to the AI platformization brought by foundation models.
Of course, this will bring new software engineering challenges, especially in responsible AI and AI services supply chain management, which are our active research agenda to expand our promptmanship and sapper IDE to address these challenges, for example, to embed our AI risk assessment framework~\cite{xia2023concrete} and \href{https://research.csiro.au/ss/science/projects/responsible-ai-pattern-catalogue/}{Responsible AI pattern catalogue} in the promptmanship, and to develop AI Bills of Materials to secure AI services supply chain~\cite{xia2023empirical}.
%This is one of our important business goals.

\section{Related Work}

\begin{figure*}
    \centering
    \includegraphics[width=\textwidth]{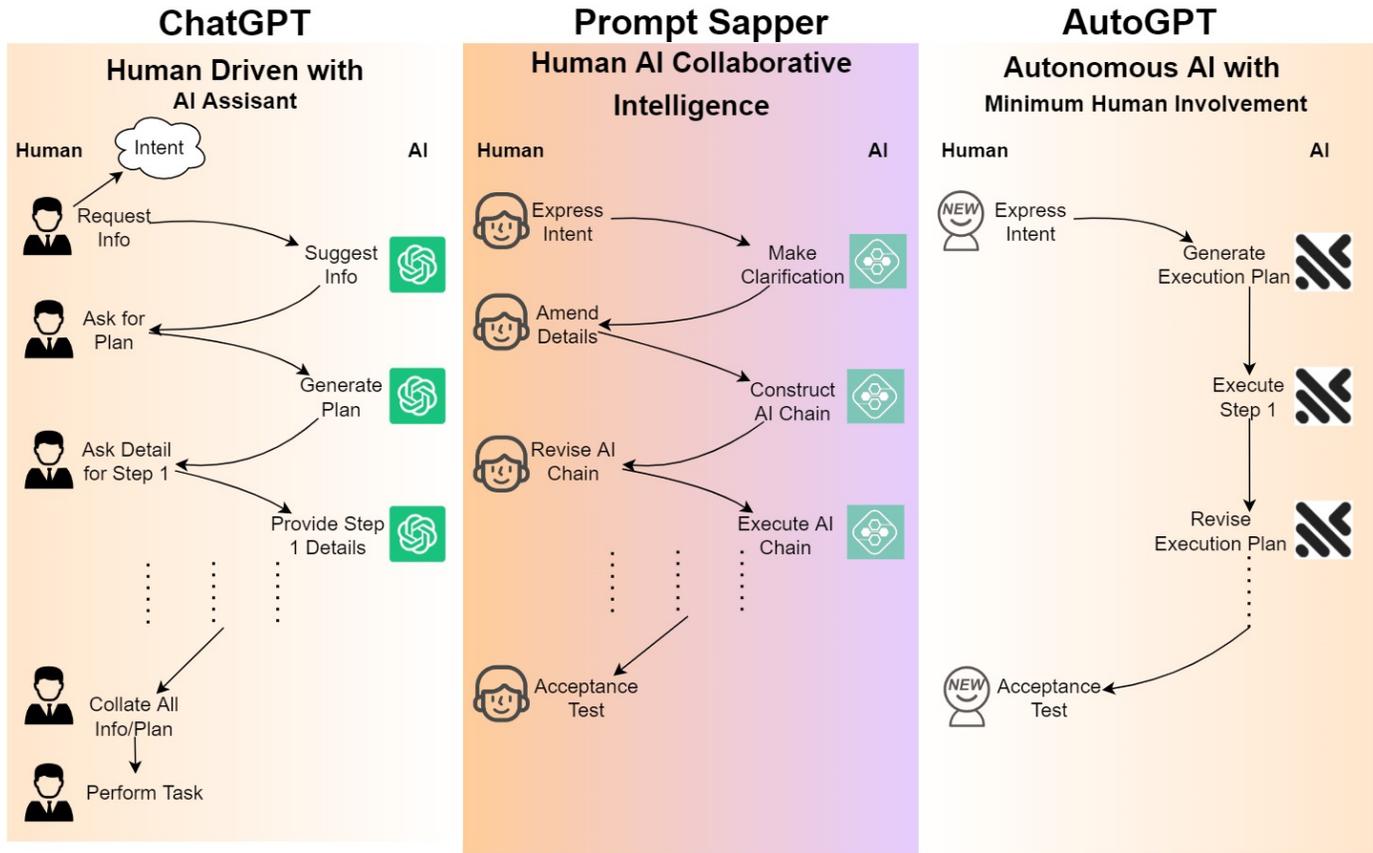}
    \caption{Spectrum of Human-AI Interaction}
    \label{fig:humanaiinteraction}
\end{figure*}

The concept of AI chain has been widely applied in various scenarios recently. 
We believe it is a fundamental strategy for determining and expanding the capability and awareness boundaries of AI.
%between the open and hidden quadrants and for expanding ability range between the open and blind quadrants, 
We categorize these works into three major categories: task-specific AI chains~\cite{arora2022ask,schick2022peer,yang2022re3,creswell2022selection,kazemi2022lambada,singh2022progprompt,huang2023pcr}, task-agnostic agent frameworks~\cite{shen2023hugginggpt, abdelghani2022gpt, wu2023visual,liang2023taskmatrix,fezarigpt,li2023camel,babyagi,hyperwriteai,transformersagents}, AI chain programming support~\cite{langchain, Dust, Primer}. 
Prompt Sapper draws inspiration from these projects and tools, but it has significant differences from them in three aspects:
%in terms of broad human-AI collaborative intelligence (as illustrated in Figure~\ref{fig:humanaiinteraction}), low requirements for computing and programming skills, and systematic framework for AI4SE4AI.

\textit{Human-AI collaborative intelligence}. 
Prompt Sapper emphasizes the collaborative interaction between AI and human users, with human focusing on requirements and acceptance while AI constructing AI chains and performing tasks.
It marries human intelligence with artificial intelligence through AI chains, effectively addressing complex problems and achieving shared goals. 
This human-AI collaborative intelligence fosters enhanced overall efficiency, reduced error rates, and empowers human users to fully harness the potential of AI. 
As illustrated in Figure~\ref{fig:humanaiinteraction}, this distinctive approach sets Prompt Sapper apart from human-driven conversational bots (e.g., ChatGPT) that demand human expertise and thought process and AI-dominated agent frameworks (e.g., AutoGPT) that rely on the LLM's divide-and-conquer and self-correction capabilities, highlighting its innovative and unique value proposition.

\textit{Requirements for programming skills}.
Compared to other projects~\cite{langchain, Dust, Primer}, Prompt Sapper significantly lowers the barrier to entry for creating complex AI services tailored to user needs. 
It introduces a suite of LLM-based virtual product manager, architect, and prompt engineer to assist users in acquiring domain knowledge, analyzing task requirements, and constructing AI chains. 
Additionally, Prompt Sapper provides an intuitive and user-friendly interface, enabling users to effortlessly interact with AI and prototype AI functionalities without the need for advanced computing or programming skills. 
This approach broadens the spectrum of people able to benefit from the advancements in AI and underscoring the distinct position of Prompt Sapper in the AI landscape.

\textit{The AI4SE4AI framework}.
Prompt Sapper values the close integration of software engineering and AI, striving to create a systematic AI4SE4AI framework. 
Within this framework, Prompt Sapper leverages AI technology to significantly improve the efficiency of software engineering processes, such as requirements analysis, AI chain design, construction, and testing. 
At the same time, Prompt Sapper adheres to and expands upon the best practices of software engineering to adapt to the new software landscape driven by AI 2.0 and Software 3.0. 
This AI4SE4AI framework not only substantially enhances the development efficiency and project quality of AI services but also supports flexible service reuse and assembly, as well as continuous improvement and optimization of AI services to meet ever-changing demands.
Low-code LLM~\cite{cai2023lowcode} and LLM pragmatism~\cite{LLMpragmatism}share the similar task decomposition idea as Prompt Sapper, but they do not provide a holistic AI4SE4AI framework.

Sapper IDE stands apart from existing Cloud IDEs or coding tools like Repit~\cite{replit}, Codespace~\cite{codespaces}, Jupyter Notebook~\cite{jupyter}. 
While these Cloud IDEs primarily address the complexity of setting up and configuring software development environments, Sapper IDE tackles the complexity of the AI-native service development process. 
Recently, several low/no-code AI tools have been introduced, such as Microsoft AI Builder~\cite{aibuilder}, Zapier~\cite{zapier}, and superbio.ai~\cite{superbioai}. 
These tools enable drag-and-drop visual programming, allowing users to utilize or customize pre-built models within their workflows. 
However, the workflows they support are often challenging to define or customize. Moreover, they primarily focus on the coding stage of the software development process (the bottom-right corner in Figure~\ref{fig:promtpmanship}). 
They do not adequately support novices in effectively expressing their needs and knowledge to AI during the early stages of the software development process, nor leveraging foundation models to build custom workflows and personalized AI services. 
Programming assistants like GitHub Co-pilot~\cite{copilot}, Replit Ghostwriter~\cite{ghostwriter}, and similar products aim to lower coding barrier and improve coding efficiency within Software 1.0/2.0. 
%These co-pilots enhance automated code generation. 
They represent useful but incremental improvements to existing IDEs (i.e., prompt to code), whereas Sapper IDE supports a disruptive transformation in software development (i.e., prompt as code).

AI chain engineering carries a certain notion of intentional programming~\cite{10.5555/345203}.
Both advocate task decomposition and intention expression.
The difference lies in that intentional programming aims to separate precise intent from source code so that software becomes easier to create and maintain in the Software 1.0/2.0 paradigms, while AI chain engineering aims to turn fuzzy informal intent into AI-native services in the era of Software 3.0.
With the support of foundation models, Software 3.0 is realizing the long-pursued vision of \href{https://www.inkandswitch.com/end-user-programming/}{end-user programming}: empowering ordinary individuals to harness the full power of computers and AI, rather than relying solely on applications developed by professional programmers.

\section{Conclusion}

We are witnessing the technological revolution in AI and software engineering. 
Prompt Sapper advocates the best practices and methodologies of AI chain engineering, driving the development and proliferation of AI-native services. 
We will adopt an ``outbound and inbound'' approach to bridge the last mile between AI-native services and end users, bringing AI chain engineering methodologies, tools, and practices to broader audience, while promoting the development of the AI service marketplace and ecosystem. 
%We believe that AI chain engineering will become one of the core technologies of the future, widely applied in various fields and industries, creating more value and benefits for humanity. 
We envision that Prompt Sapper on top of foundation models will fulfil the vision of personal AI, and propel society and the economy towards a more intelligent future.
%helping us solve problems more rapidly, improve work efficiency, 
% or


\begin{comment}
\appendices
\section{Proof of the First Zonklar Equation}
Appendix one text goes here.

\section{}
Appendix two text goes here.
\end{comment}

\ifCLASSOPTIONcompsoc
  \section*{Acknowledgments}
\else
  \section*{Acknowledgment}
\fi

The authors would like to thank...

\ifCLASSOPTIONcaptionsoff
  \newpage
\fi



\bibliographystyle{IEEEtran}
% \bibliography{IEEEabrv,../bib/paper}
\bibliography{reference.bib}


% \end{thebibliography}


\begin{comment}
\begin{IEEEbiography}{Michael Shell}
Biography text here.
\end{IEEEbiography}

\begin{IEEEbiographynophoto}{John Doe}
Biography text here.
\end{IEEEbiographynophoto}

%\newpage

\begin{IEEEbiographynophoto}{Jane Doe}
Biography text here.
\end{IEEEbiographynophoto}


%\vfill

\end{comment}

\end{document}